\documentclass[12pt]{article}
\usepackage{color}
\usepackage{amssymb,amsmath}
\usepackage[dvips]{graphicx}
\newcommand{\bea}{\begin{eqnarray}}
\newcommand{\eea}{\end{eqnarray}}
\begin{document}
\setlength{\baselineskip}{0.7cm}
\setlength{\baselineskip}{0.7cm}
\begin{titlepage} 
\begin{flushright}
OCU-PHYS-390
\end{flushright}
\vspace*{10mm}
\begin{center}{\Large\bf 125 GeV Higgs Boson and TeV Scale Colored \\ 
\vspace*{2mm}
Fermions 
in Gauge-Higgs Unification}
\end{center}
\vspace*{10mm}
\begin{center}
{\large Nobuhito Maru}$^{a}$ 
and 
{\large Nobuchika Okada}$^{b}$
\end{center}
\vspace*{0.2cm}
\begin{center}
${}^{a}${\it Department of Mathematics and Physics, Osaka City University, \\
Osaka 558-8585, Japan}
\\[0.2cm]
${}^{b}${\it Department of Physics and Astronomy, University of Alabama, \\
Tuscaloosa, Alabama 35487, USA} 
\end{center}
\vspace*{2cm}
\begin{abstract} 

In the context of a simple gauge-Higgs unification scenario  
 based on the gauge group $SU(3)\times U(1)'$ 
 in a 5-dimensional flat spacetime, 
 we investigate a possibility to realize the Higgs boson mass
 of 125 GeV with a compactification scale at the TeV. 
With the introduction of colored bulk fermions in a certain representation 
 under the gauge group with a half-periodic boundary condition, 
 we analyze the one-loop RGE of the Higgs quartic coupling 
 with the gauge-Higgs condition and successfully obtain 
 125 GeV Higgs boson mass by adjusting a bulk mass 
 for the fermions with a fixed compactification scale. 
The Kaluza-Klein modes of the colored fermions 
 contribute to the Higgs-to-digluon effective coupling 
 through the one-loop corrections. 
Since the contribution is destructive to the Standard Model one 
 and reduces the Higgs boson production cross section, 
 the recent LHC result gives a lower bound 
 on the mass of the lightest Kaluza-Klein mode fermion. 
We find the mass of the lightest Kaluza-Klein mode fermion
 in the range of $2-3$ TeV for $10\%-5\%$ reductions 
 of the Higgs boson production cross section 
 through the gluon fusion channel. 
Such Kaluza-Klein mode fermions can be discovered 
 at the LHC run II with $\sqrt{s}=13-14$ TeV, 
 and LHC phenomenology for the Kaluza-Klein fermions 
 is briefly discussed. 

\end{abstract}
\end{titlepage}

\section{Introduction}

The discovery of the long-sought Higgs boson 
 by ATLAS~\cite{ATLAS} and CMS~\cite{CMS} collaborations 
 at the Large Hadron Collider (LHC) 
 is a milestone in the history of particle physics. 
With the discovery, we have begun testing the Higgs sector 
 of the Standard Model (SM). 
Although the observed data for a variety of Higgs boson 
 decay modes are found to be mostly consistent with the SM expectations, 
 we need more data for definite conclusions. 
In fact, the Higgs diphoton decay mode showed 
 the signal strength considerably larger than the SM prediction. 
Since the Higgs-to-diphoton coupling arises at the quantum level 
 even in the SM, there is a good chance that the deviation 
 originates from a certain new physics effect. 
This has motivated many recent studies for explanation 
 of the deviation in various extensions of the SM 
 with supersymmetry~\cite{AHC-SUSY} or 
 without supersymmetry~\cite{AHC-NonSUSY}. 
The excess persists in the updated ATLAS analysis~\cite{ATLAS2}, 
 while the updated CMS analysis~\cite{CMS2} gives a much lower value 
 for the signal strength of the diphoton events 
 than the previous one~\cite{CMS}. 
More precise measurements of the Higgs boson properties 
 might be a clue to finding physics beyond the SM, 
 although the current LHC data have no indication for it.

In this paper we consider the gauge-Higgs unification (GHU) 
 scenario \cite{GH} as a candidate for physics beyond the SM. 
This scenario offers a solution to the gauge  hierarchy problem 
 without invoking supersymmetry, where the SM Higgs doublet 
 is identified as an extra spatial component of the gauge field 
 in higher dimensional theory. 
The scenario predicts various finite physical observables 
 such as Higgs potential~\cite{1loopmass, 2loop}, 
 the partial decay widths of the Higgs boson 
 to digluon and diphoton~\cite{MO, Maru, MO2}, 
 the anomalous magnetic moment $g-2$~\cite{g-2}, 
 and the electric dipole moment~\cite{EDM}, 
 thanks to the higher dimensional gauge symmetry, 
 irrespective of the non-renormalizable theory.

In our previous work \cite{MO2, MO3},  
 we have investigated a simple GHU model 
 based on the $SU(3) \times U(1)'$ gauge group 
 by introducing color-singlet bulk fermions 
 in 10 or 15-representation under the $SU(3)$ 
 with a half-periodic boundary condition.  
We have analyzed one-loop contributions of 
 Kaluza-Klein (KK) modes to the Higgs-to-digluon 
 and Higgs-to-diphoton couplings 
 and have confirmed that the KK mode contributions 
 are destructive to the SM contributions 
 from corresponding SM particles in the model \cite{MO}. 
This is a remarkable feature of the GHU, 
 closely related to the absence of the quadratic divergence 
 in Higgs self-energy corrections. 
We have shown that the bulk fermions can enhance 
 the Higgs-to-diphoton coupling with appropriately chosen 
 electric charges \cite{MO2}. 
The bulk fermions also play a crucial role 
 to achieve a Higgs boson mass of around 125 GeV. 
Without the bulk fermions, Higgs boson mass is predicted 
 to be too small, less than 100 GeV.

In the same context, we have also studied 
 the KK mode contributions to the Higgs to $Z \gamma$ decay \cite{MO3}. 
We have found a very striking result that 
 the KK mode contributions at the 1-loop level do not exist 
 and hence the effective Higgs to $Z \gamma$ coupling 
 remains unchanged, irrespectively of 
 the effective Higgs-to-diphoton coupling. 
This is because $Z$ boson only couples 
 with two different mass eigenstates 
 while the Higgs boson and photon couple with the same mass eigenstates. 
As a result, there is no one-loop diagram with KK modes 
 contributing to the effective Higgs-to-$Z \gamma$ coupling. 
This specific result originates from the basic structure 
 of the GHU scenario, where the SM gauge group is embedded 
 in a larger gauge group in extra-dimensions and 
 the SM Higgs doublet is identified with 
 the higher dimensional component of the bulk gauge field.

In this paper, we investigate the case that 
 the new bulk fermions with the half periodic boundary condition 
 are color triplets, instead of color singlet ones 
 introduced in \cite{MO2}. 
As has been studied in \cite{MO2},  
 we can realize the 125 GeV Higgs boson mass 
 also in this case by adjusting the bulk mass 
 for a fixed compactification scale, 
 depending on the representation of the bulk fermions. 
However, since the colored KK modes 
 contribute to the effective Higgs-to-digluon coupling 
 destructively to the SM particle (top quark) contribution, 
 the Higgs boson production cross section 
 through the gluon fusion is dramatically reduced 
 as the KK modes become light. 
As a result, the current LHC data for the Higgs boson production 
 give the lower bound on the lightest KK mode mass. 
For a variety of representations under the $SU(3)\times U(1)'$ 
 gauge group for the bulk fermions, 
 we identify the model-parameter region 
 which can reproduce the 125 GeV Higgs boson mass 
 and satisfy the LHC constraint.

The plan of this paper is as follows. 
In the next section, we introduce a simple GHU model \cite{SSS, CCP} 
 based on the gauge group $SU(3)\times U(1)'$ 
 in a 5-dimensional flat spacetime 
 with an orbifold $S^1/Z_2$ compactification 
 to the 5th spacial dimension. 
We consider the introduction of a variety of colored bulk fermions 
 in the representations of ${\bf 6}, {\bf 8}, {\bf 10}$ 
 and ${\bf 15}$ under the bulk $SU(3)$ gauge group, 
 for which a half-periodic boundary condition is imposed. 
In this context, we estimate Higgs boson mass 
 in a four dimensional effective theory approach developed 
 in Ref.~\cite{GHcondition}, where Higgs boson mass 
 is calculated by solving 1-loop renormalization group equation (RGE) 
 of the Higgs quartic coupling with 
 the so-called gauge-Higgs condition \cite{GHcondition}. 
Once the representation of the bulk fermions are fixed, 
 we identify the model-parameter region so as to 
 reproduce the 125 GeV Higgs boson mass. 
In Sec.~3, we study effects of the colored bulk fermions 
 to Higgs-to-digluon and Higgs-to-diphoton couplings. 
We discuss the lower bound on the mass of the lightest KK mode 
 fermions to be consistent with the LHC data 
 for the Higgs boson signals. 
Sec.~4 is devoted to conclusions and discussions.

\section{Higgs boson mass in simple GHU model}

Let us consider a simple GHU model based on the gauge group 
 $SU(3) \times U(1)'$ in a 5-dimensional flat spacetime 
 with an orbifolding of the 5th dimension 
 on $S^1/Z_2$ with radius $R_c$ of $S^1$. 
In our setup of bulk fermions 
 whose zero-modes correspond to the SM fermions, 
 we follow Ref.~\cite{CCP}: 
 the up-type quarks except for the top quark, 
 the down-type quarks and the leptons are embedded,  
 respectively, into ${\bf 3}$, $\overline{{\bf 6}}$, 
 and ${\bf 10}$ representations of $SU(3)$. 
In order to realize the large top Yukawa coupling, 
 the top quark is embedded into a rank $4$ representation 
 of $SU(3)$, namely $\overline{{\bf 15}}$. 
The extra $U(1)'$ symmetry works to yield 
 the correct weak mixing angle, and the SM $U(1)_Y$ gauge boson 
 is given by a linear combination between the gauge bosons 
 of the $U(1)'$ and the $U(1)$ subgroup in $SU(3)$ \cite{SSS}\footnote{
 It is known that the correct Weinberg angle can be also obtained 
 by introducing brane localized gauge kinetic terms \cite{SSS}, 
 but we do not take this approach in this paper.}.  
We assign appropriate $U(1)'$ charges for bulk fermions 
 to give the correct hypercharge for the SM fermions.

The boundary conditions should be suitably assigned 
 to reproduce the SM fields as the zero modes. 
While a periodic boundary condition corresponding to $S^1$ 
 is taken for all of the bulk SM fields, 
 the $Z_2$ parity is assigned for gauge fields and fermions 
 in the representation ${\cal R}$ 
 by using the parity matrix $P={\rm diag}(-,-,+)$ as
\bea
A_\mu (-y) = P^\dag A_\mu(y) P, \quad A_y(-y) =- P^\dag A_\mu(y) P,  \quad 
\psi(-y) = {\cal R}(P)\psi(y) 
\label{parity}
\eea 
 where the subscripts $\mu$ ($y$) denotes the four (the fifth) 
 dimensional component. 
With this choice of parities, the $SU(3)$ gauge symmetry is 
 explicitly broken to $SU(2) \times U(1)$. 
A hypercharge is a linear combination of $U(1)$ and $U(1)'$ 
 in this setup. 
Here, the $U(1)_X$ symmetry is anomalous in general and 
 broken at the cutoff scale and hence,  
 the $U(1)_X$ gauge boson has a mass of order 
 of the cutoff scale \cite{SSS}. 
As a result, zero-mode vector bosons in the model are 
 only the SM gauge fields.

Off-diagonal blocks in $A_y$ have zero modes 
 because of the overall sign in Eq.~(\ref{parity}), 
 which corresponds to an $SU(2)$ doublet. 
In fact,  the SM Higgs doublet ($H$) is identified as 
\bea
A_y^{(0)} = \frac{1}{\sqrt{2}}
\left(
\begin{array}{cc}
0 & H \\
H^\dag & 0 \\
\end{array}
\right). 
\eea
The KK modes of $A_y$ are eaten by KK modes of the SM gauge bosons 
 and enjoy their longitudinal degrees of freedom 
 like the usual Higgs mechanism.

The parity assignment also provides the SM fermions as massless modes, 
 but it also leaves exotic fermions massless. 
Such exotic fermions are made massive by introducing 
 brane localized fermions with conjugate $SU(2) \times U(1)$ charges 
 and an opposite chirality to the exotic fermions, 
 allowing us to write mass terms on the orbifold fixed points. 
In the GHU scenario, the Yukawa interaction is unified 
 with the gauge interaction, so that the SM fermions 
 obtain the mass of the order of the $W$-boson mass 
 after the electroweak symmetry breaking. 
To realize light SM fermion masses, one may introduce 
 a $Z_2$-parity odd bulk mass terms for the SM fermions, 
 except for the top quark. 
Then, zero mode fermion wave functions with opposite chirality 
 are localized towards the opposite orbifold fixed points 
 and as a result, their Yukawa coupling is exponentially 
 suppressed by the overlap integral of the wave functions. 
In this way, all exotic fermion zero modes become heavy 
 and small Yukawa couplings for light SM fermions 
 are realized by adjusting the bulk mass parameters. 
In order to realize the top quark Yukawa coupling, 
 we introduce a rank $4$ symmetric tensor 
 $\overline{{\bf 15}}$ without a bulk mass \cite{CCP}. 
This provide us with a group theoretical factor $2$ 
 for the coupling of the top quark with the Higgs doublet, 
 so that we have $m_t = 2 m_W$ at the compactification scale \cite{SSS}. 
Taking QCD threshold corrections to the top quark pole mass, 
 this mass relation is desirable. 
See, for example, Ref.~\cite{FlavorCP} 
 for flavor mixing and CP violation in the GHU scenario.

Since it is a highly non-trivial task to propose 
 a realistic GHU scenario (see, for example, 
 Refs.~\cite{SSS, CCP} toward this direction), 
 we employ a four dimensional effective theory approach 
 developed by Ref.~\cite{GHcondition} 
 in estimating a Higgs boson mass. 
As has been shown in Ref.~\cite{GHcondition}, 
 the low energy effective theory of the 5-dimensional GHU scenario 
 is equivalent to the SM with the so-called ``gauge-Higgs condition'' 
 on the Higgs quartic coupling, 
 namely, we impose a vanishing Higgs quartic coupling
 at the compactification scale. 
This boundary condition reflects the 5-dimensional gauge 
 invariance and, once it is restored, the Higgs potential disappears. 
According to this effective theory approach, 
 the Higgs boson mass at low energies is easily calculated 
 from the RGE evolution of the Higgs quartic coupling 
 below the compactification scale,  
 under the the gauge-Higgs condition. 
Another advantage of this approach is that 
 we can see that the GHU scenario offers a natural solution 
 to the instability problem of the Higgs potential in the SM, 
 because the potential vanishes at the compactification scale 
 and remains zero at higher scales. 
We assume that the electroweak symmetry breaking correctly occurs  
 by the introduction of a suitable set of bulk fermions. 
Note that the effective Higgs mass squared is quadratically 
 sensitive to the mass of heavy states, 
 while the effective Higgs quartic coupling is dominantly  
 determined by interactions of the Higgs doublet with light states. 
Therefore, the Higgs boson mass at low energies 
 is mainly determined by light states below the compactification scale, 
 assuming the correct electroweak symmetry breaking.

In the following, we consider three different representations 
 for the colored bulk fermions: $({\bf 3}, {\bf 6})$, 
 $({\bf 3}, {\bf 10})$ and $({\bf 3}, {\bf 15})$ 
 under $SU(3)_c \times SU(3)$, 
 with suitable $U(1)'$ charge assignments. 
The half periodic boundary condition is imposed 
 on the bulk fermions, $\psi(y+2\pi R) = -\psi(y)$. 
Thanks to the boundary condition, 
 the model leads to no unwanted exotic massless fermion, 
 while the lightest KK modes of the fermions 
 are involved in the RGE evolutions 
 since their masses are smaller than the compactification scale. 
We will see that the existence of the half-periodic bulk fermion 
 is crucial to achieve a Higgs boson mass of around 125 GeV

Let us begin with the ${\bf 6}$-plet of $SU(3)$, 
 which is decomposed into the representations
 under $SU(2) \times U(1)$ as  
\bea
 {\bf 6} = {\bf 1}_{-2/3} \oplus {\bf 2}_{-1/6} \oplus {\bf 3}_{1/3},  
\label{6deco}
\eea
 where the numbers in the subscript denote the $U(1)$ charges. 
After the electroweak symmetry breaking 
 the KK mass spectrum is found as follows: 
\bea
&& 
\left( m_{n,-2/3}^{(\pm)} \right)^2 
 = \left( m_{n+\frac{1}{2}} \pm 2 m_W \right)^2 +M^2,~~ 
   m_{n+\frac{1}{2}}^2 + M^2, \nonumber \\ 
&& 
\left( m_{n,+1/3}^{(\pm)} \right)^2 
 = \left( m_{n+\frac{1}{2}} \pm m_W \right)^2 +M^2, \nonumber \\ 
&& 
\left( m_{n,+4/3}^{(\pm)} \right)^2 
 = m_{n+\frac{1}{2}}^2 +M^2, 
\label{6KKspectrum}
\eea  
 where the numbers in the subscript denote 
 the ``electric charges"\footnote{
 Here ``electric charges" mean by electric charges 
 of $SU(2) \times U(1) \subset SU(3)$. 
 A true electric charge of each KK mode is given 
 by a sum of the ``electric charge" and $U(1)'$ charge $Q$.
 }  
 of the corresponding KK mode  fermions, 
 $m_{n+\frac{1}{2}}= \left( n+\frac{1}{2}\right) M_{\rm KK}$ 
 with $n=0,1,2,\cdots$, $M_{\rm KK} \equiv 1/R_c$ and $M$ is a bulk mass. 
In the same way, we decompose the ${\bf 10}$-plet as 
\bea
 {\bf 10} = {\bf 1}_{-1} \oplus {\bf 2}_{-1/2} 
 \oplus {\bf 3}_{0} \oplus {\bf 4}_{1/2}.   
\label{10deco}
\eea
The KK mass spectrum after the electroweak symmetry breaking is found as: 
\bea
&& 
\left( m_{n,-1}^{(\pm)} \right)^2 
 = \left( m_{n+\frac{1}{2}} \pm 3 m_W \right)^2 +M^2,~~ 
   \left( m_{n+\frac{1}{2}} \pm   m_W \right)^2 +M^2, \nonumber \\ 
&& 
\left( m_{n,0}^{(\pm)} \right)^2 
 = \left( m_{n+\frac{1}{2}} \pm 2 m_W \right)^2 +M^2,~~ 
   m_{n+\frac{1}{2}}^2 + M^2, \nonumber \\ 
&& 
\left( m_{n,+1}^{(\pm)} \right)^2 
 = \left( m_{n+\frac{1}{2}} \pm m_W \right)^2 +M^2, \nonumber \\ 
&& 
\left( m_{n,+2}^{(\pm)} \right)^2 
 = m_{n+\frac{1}{2}}^2 + M^2. 
\label{10KKspectrum}
\eea  
For the ${\bf 15}$-plet, the decomposition 
 under $SU(2) \times U(1)$ is given by 
\bea
 {\bf 15} = {\bf 1}_{-4/3} \oplus {\bf 2}_{-5/6} 
 \oplus {\bf 3}_{-1/3} \oplus {\bf 4}_{1/6} 
 \oplus {\bf 5}_{2/3}.  
\label{15deco}
\eea
After the electroweak symmetry breaking, 
 the KK mass spectrum is found as follows: 
\bea
&& 
\left( m_{n,-4/3}^{(\pm)} \right)^2 
 = \left( m_{n+\frac{1}{2}} \pm 4 m_W \right)^2 +M^2,~~ 
   \left( m_{n+\frac{1}{2}} \pm 2  m_W \right)^2 +M^2,~~  
   m_{n+\frac{1}{2}}^2+M^2, 
  \nonumber \\ 
&& 
\left( m_{n,-1/3}^{(\pm)} \right)^2 
 = \left( m_{n+\frac{1}{2}} \pm 3 m_W \right)^2 +M^2,~~~ 
   \left( m_{n+\frac{1}{2}} \pm   m_W \right)^2 +M^2, 
  \nonumber \\ 
&& 
\left( m_{n,2/3}^{(\pm)} \right)^2 
 = \left( m_{n+\frac{1}{2}} \pm 2 m_W \right)^2 +M^2,~~ 
   m_{n+\frac{1}{2}}^2+M^2, 
  \nonumber \\ 
&&
\left( m_{n,5/3}^{(\pm)} \right)^2 
 = \left( m_{n+\frac{1}{2}} \pm m_W \right)^2 +M^2,
  \nonumber \\ 
&&
\left( m_{n,8/3}^{(\pm)} \right)^2 
 = m_{n+\frac{1}{2}}^2+M^2. 
\label{15KKspectrum}
\eea  

In our model, we have introduced bulk fermions 
 with the half-periodic boundary condition, 
 and their first KK modes appear 
 below the compactification scale. 
Therefore, not only the SM particles but also 
 the first KK modes are involved in our RGE analysis 
 with the gauge-Higgs condition.\footnote{
In Ref.~\cite{GHUcond-mh}, the gauge-Higgs condition
 with only the SM particle contents below the compactification scale 
 is used to predict Higgs boson mass as a function of 
 the compactification scale.
}
The 1-loop RGE for the Higgs quartic coupling $\lambda$ 
 below the compactification scale is given by 
\bea
\frac{d \lambda}{d \ln \mu} &=& 
\frac{1}{16\pi^2} 
\left[
12 \lambda^2 
- \left( \frac{9}{5} g_1^2 + 9 g_2^2 \right) \lambda
+ \frac{9}{4} \left( \frac{3}{25}g_1^4 + \frac{2}{5}g_1^2 g_2^2 + g_2^4 \right) \right. \nonumber \\
&& \left. + 4 \left( 
  3 y_t^2 + 3 C_2({\bf R}) \left( \frac{g_2}{\sqrt{2}}\right)^2 
 \right) \lambda 
 -4 \left(3 y_t^4 + 3 C_4({\bf R})
 \left( \frac{g_2}{\sqrt{2}}\right)^4 
 \right)  \right], 
\label{RGEup}
\eea
 where $y_t$ is the top Yukawa coupling, 
 $g_{1,2}$ are the $SU(2)$, $U(1)_Y$ gauge couplings, respectively,
 and $C_2({\bf R})$  and $C_4({\bf R})$ are 
 contributions to the beta function 
 by the representation ${\bf R}={\bf 6}$, ${\bf 10}$, and ${\bf 15}$. 
In our RGE analysis, we neglect the KK mode mass splitting 
 by the electroweak symmetry breaking 
 and set the first KK mode mass as 
\bea 
 m_0^{(\pm)}= \frac{1}{2} M_{\rm KK} \sqrt{1+4 c_B^2}
\eea 
where $c_B \equiv M/M_{\rm KK}$.

For the energy scale $m_0^{(\pm)} \leq \mu \leq M_{\rm KK}$, 
 the coefficients $C_2({\bf R})$ and $C_4({\bf R})$ 
 are explicitly given by 
\bea
C_2({\bf 6}) &=& 2 \left( 2^2 + 1^2 \right), 
\nonumber \\ 
C_4({\bf 6}) &=& 2 \left( 2^4 + 1^4 \right), 
\nonumber \\ 
C_2({\bf 10}) &=& 2 \left( 3^2 + 1^2 + 2^2 + 1^2 \right), 
\nonumber \\ 
C_4({\bf 10}) &=& 2 \left( 3^4 + 1^4 + 2^4 + 1^4 \right), 
\nonumber \\ 
C_2({\bf 15}) &=& 2 \left( 4^2 + 2^2 + 3^2 + 1^2 + 2^2 + 1^2 \right), 
\nonumber \\ 
C_4({\bf 15}) &=& 2 \left( 4^4 + 2^4 + 3^4 + 1^4 + 2^4 + 1^4 \right), 
\eea
 while these coefficients are set to be $0$ for $\mu < m_0^{(\pm)}$ 
 since the first KK modes are decoupled at the scale $m_0^{(\pm)}$. 
In our analysis, the running effects 
 for $y_t, g_{1,2}$ are simply neglected.

\begin{figure}[ht]
  \begin{center}
 \includegraphics[width=130mm]{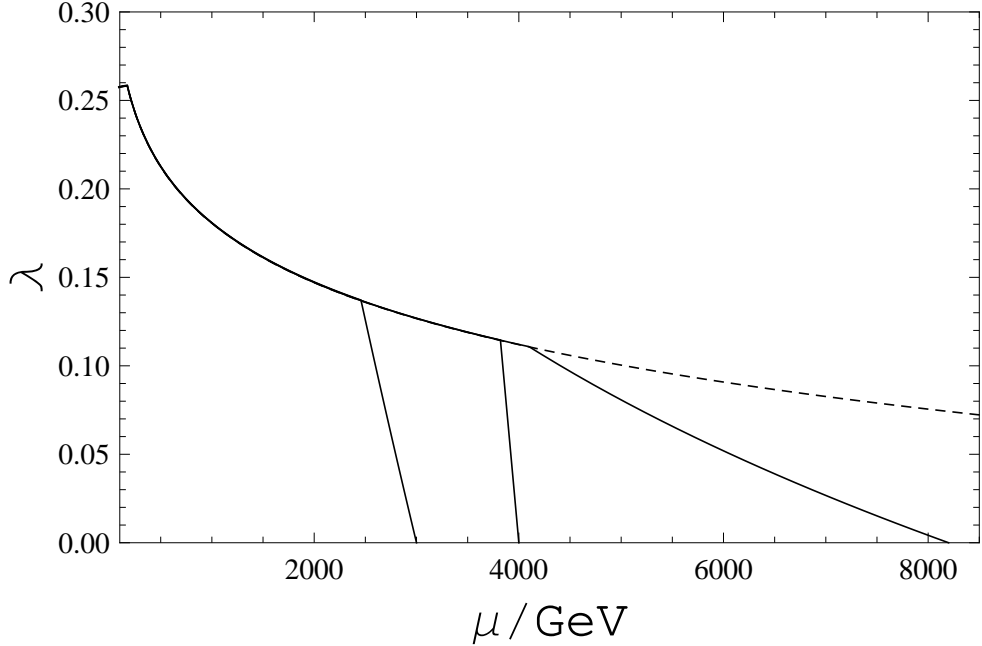}
   \end{center}
  \caption{
The RGE running of the Higgs quartic coupling at the 1-loop level. 
The three solid lines correspond, respectively,
 to the cases of 
 the ${\bf 6}$-plet 
 with $(M_{\rm KK}, m_0^{(\pm)})=(8.2, 4.1)$ TeV,   
 the ${\bf 10}$-plet 
 with $(M_{\rm KK}, m_0^{(\pm)})=(3.0, 2.46)$ TeV,   
 the ${\bf 15}$-plet 
 with $(M_{\rm KK}, m_0^{(\pm)})=(4.0, 3.82)$ TeV. 
The dashed line shows the RGE running of 
 the SM Higgs quartic coupling with the boundary condition,
 $\lambda(\mu=m_h)=0.258$, 
 corresponding to the Higgs pole mass $m_h=125$ GeV. 
}
  \label{RGElambda}
\end{figure}

The numerical solution to the 1-loop RGE of the Higgs quartic coupling 
 are shown in Fig.~\ref{RGElambda}. 
Here we have applied the gauge-Higgs condition 
 at the compactification scales, 
 $M_{\rm KK}=8.2$, $3.5$, and $4.0$ TeV, 
 respectively for ${\bf 6}$, ${\bf 10}$, and ${\bf 15}$-plets 
 and numerically solve the RGE of Eq.~(\ref{RGEup}) toward low energies. 
Here, we have used 
 $y_t(\mu) =0.944$ for $\mu \geq m_t=173.2$ GeV 
 ($y_t(\mu)=0$ for $\mu < m_t=173.2$ GeV), 
 and $g_1=0.459$, and $g_2=0.649$ at the $Z$-boson mass scale. 
For simplicity, we estimate the Higgs boson pole mass 
 by the condition $ \lambda(\mu=m_h) v^2 =m_h^2$, 
 which means $\lambda (\mu=m_h) = 0.258$ 
 for $m_h=125$ GeV. 
For a fixed compactification scale, 
 we have adjusted the bulk mass $M$ 
 so as to reproduce the 125 GeV Higgs boson mass. 
Corresponding masses for the fist KK modes are 
 found to be $m_0^{(\pm)} = 4.1, 2.46$ and $3.82$ TeV, respectively, 
 for ${\bf 6}$, ${\bf 10}$, and ${\bf 15}$-plets. 
In this rough analysis, the Higgs quartic coupling 
 under the SM RGE evolution becomes zero 
 at $\mu \simeq 3 \times 10^4$ GeV. 
As is well-known, in more precise analysis with higher 
 order corrections (see, for example, \cite{RGE-Higgs}), 
 the Higgs quartic coupling becomes zero 
 at $\mu \simeq 10^{10}$ GeV. 
In the precise analysis, the running top Yukawa coupling 
 is monotonically decreasing and the higher order corrections 
 positively contribute to the beta function, 
 as a result, the scale realizing $\lambda(\mu)=0$ 
 is pushed up to high energies.

As can be seen from Fig.~\ref{RGElambda},
 the existence of the half-periodic bulk fermions 
 is essential to realize the 125 GeV Higgs boson mass  
 with the compactification scale at the TeV. 
For a fixed compactification scale,  
 we adjust the bulk mass to be the value 
 at which the RGE solution with the bulk fermions 
 merges with the SM RGE evolution toward high energies 
 with the boundary condition $\lambda (\mu=m_h) = 0.258$. 
Since a higher dimensional bulk fermions provide 
 more KK mode fermions in the SM decomposition, 
 the running of Higgs quartic coupling is more sharply 
 rising from zero toward low energies 
 in the case with fermions in higher representations. 
In the viewpoint of the LHC physics, 
 the ${\bf 10}$ and ${\bf 15}$-plets are interesting 
 since their first KK mode masses can be around a few TeV, 
 while for the ${\bf 6}$-plet the KK mode mass 
 is relatively higher. 
This is because the ${\bf 6}$-plet provides less number 
 of the lightest KK mode fermions and a relatively larger 
 compactification scale is required to reproduce 
 the 125 GeV Higgs boson mass. 
In fact, we have found the lower bound, $M_{\rm KK} \geq 8.2$ TeV,
 for the case with the ${\bf 6}$-plet. 
We have also performed the same analysis for a bulk fermion 
 of the ${\bf 8}$-plet and found the result very similar
 to the ${\bf 6}$-plet case. 
Hereafter we will focus on two cases of 
 ${\bf 10}$ and ${\bf 15}$-plets.

\begin{figure}[htbp]
  \begin{center}
   \includegraphics[width=130mm]{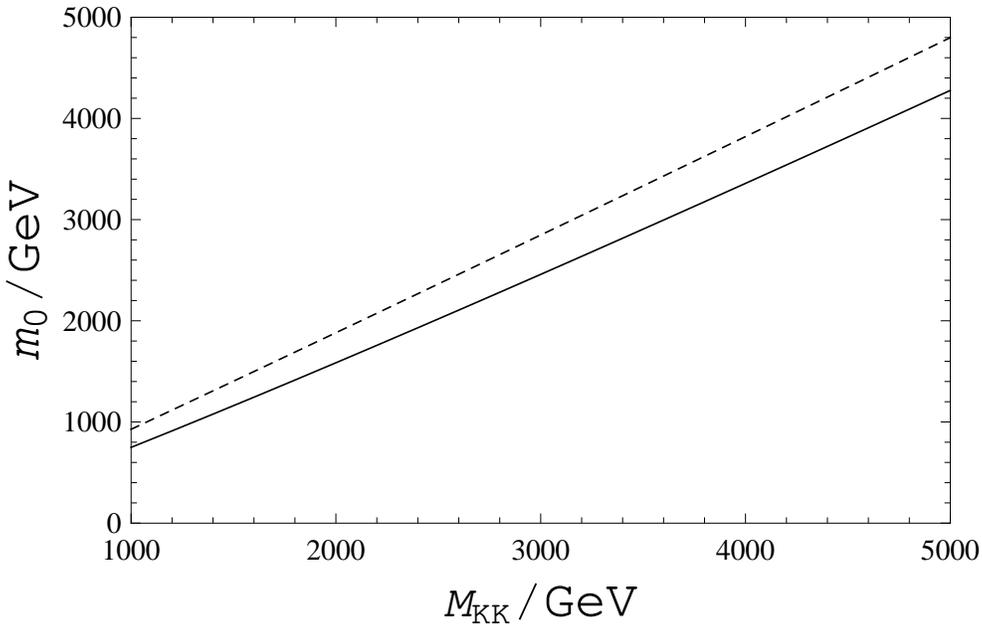}
   \end{center}
  \caption{
For the case of the ${\bf 10}$-plet (${\bf 15}$-plet), 
 the solid (dashed) line shows the relation between 
 the lightest KK mode mass $m_0^{(\pm)}$ 
 and the compactification scale $M_{\rm KK}$, 
 along which $m_h=125$ GeV is reproduced.
}
  \label{M0vsMKK}
\end{figure}

Once the compactification scale is fixed,  
 the bulk mass parameter ($c_B$) or equivalently, 
 the lightest KK mode mass ($m_0^{(\pm)}$) 
 is determined so as to reproduce the 125 GeV Higgs boson mass. 
The relation between $m_0^{(\pm)}$ and $M_{\rm KK}$
 is shown in Fig. \ref{M0vsMKK}. 
The solid (dashed) line represents the resultant bulk mass 
 $m_0^{(\pm)}$ of the ${\bf 10}({\bf 15})$-plet 
 as a function of the compactification scale, $M_{\rm KK}$. 
We have found that the lightest KK mode mass $m_0^{(\pm)}$ 
 are close to the compactification scale $M_{\rm KK}$.

\section{Higgs production and diphoton decay in GHU}
The bulk colored fermions we have introduced 
 contribute to the effective Higgs-to-digluon 
 and diphoton couplings 
 through 1-loop corrections~\cite{MO, MO2}, 
 so that the Higgs boson production cross section 
 and its branching ratio can be altered from 
 the SM values. 
In this section, we evaluate the contributions 
 from  the bulk ${\bf 10}$-plet and ${\bf 15}$-plet fermions
 and discuss constraints for the model-parameters 
 by the LHC data.

\subsection{Higgs boson production through gluon fusion at the LHC} 
At the LHC, the Higgs boson is dominantly produced  
 via gluon fusion process 
 with the following dimension five operator 
 between the Higgs boson and digluon: 
\bea
{\cal L}_{{\rm eff}} = C_{gg} h G^a_{\mu\nu} G^{a\mu\nu} 
\label{dim5g}
\eea
 where $h$ is the SM Higgs boson, and $G^a_{\mu\nu}$ 
 ($a=1$-$8$) is the gluon field strength.   
The SM contribution to $C_{gg}$ is dominated 
 by top quark 1-loop corrections. 
As a good approximation, we express the contribution 
 by using the Higgs low energy theorem \cite{HLET}, 
\bea
C_{gg}^{{\rm SM top}} \simeq \frac{g_3^2}{32\pi^2 v} 
 b_3^t \frac{\partial}{\partial \log v} \log m_t 
 = \frac{\alpha_s}{12\pi v}
\label{SMg}
\eea
 where $g_3$ ($\alpha_s=g_3^2/(4 \pi))$ is 
 the QCD coupling constant (fine structure constant), 
 $m_t$ is a top quark mass, 
 and $b_3^t=2/3$ is a top quark contribution 
 to the beta function coefficient of QCD.

In addition to the SM contribution, 
 we take into account the contributions from 
 the top quark KK-modes and the KK modes of 
 the bulk ${\bf 10}$-plet or ${\bf 15}$-plet.  
As mentioned before, the SM top quark is embedded 
 into the $\overline{{\bf 15}}$-plet 
 with a periodic boundary condition 
 and its KK mass spectrum is given by~\cite{CCP}
\bea
 m_{n,t}^{(\pm)} = m_n \pm m_t 
 \label{KKtopmass}
\eea
 where $m_t=2 m_W$ with the $W$-boson mass ($m_W =80.4$ GeV), 
 and $m_n \equiv n M_{\rm KK}$ with an integer $n=1,2,3,\cdots$. 
Although the $\overline{{\bf 15}}$-plet generally includes 
 exotic massless fermions, we have assumed that 
 all the exotic fermions are decoupled 
 by adjusting large brane-localized mass terms 
 with the brane fermions. 
Thus, we only consider KK modes of the SM top quark.\footnote{
 One might think that the KK mode contributions from the light 
 fermions should be taken into account. 
 However, they can be safely neglected compared to those  
 from the heavy fermions, since the total KK mode sum is 
 proportional to the corresponding SM fermion masses generated by 
 the electroweak symmetry breaking as is seen in (\ref{KKgfusion}) 
 for instance.} 
It is straightforward to calculate KK top contributions 
 by using the Higgs low energy theorem: 
\bea
C_{gg}^{{\rm KKtop}} &\simeq& \frac{\alpha_s}{12\pi v}  
 \sum_{n=1}^\infty \frac{\partial}{\partial \log v} 
 \left[ \log (m_n + m_t) + \log(m_n - m_t) \right] \nonumber \\
&\simeq&  - \frac{\alpha_s}{12\pi v} 2 \sum_{n=1}^\infty 
 \left( \frac{2 m_W}{m_n} \right)^2
=
- \frac{\alpha_s}{12\pi v} \times \frac{\pi^2}{3} \left( \frac{2 m_W}{M_{\rm KK}} \right)^2, 
\label{KKgfusion}
\eea
 where we have used $m_t=2 m_W$, 
 an approximation of $m_W^2 \ll m_n^2$, 
 and $\sum_{n=1}^\infty 1/n^2 = \pi^2/6$.

The ${\bf 10}$-plet or ${\bf 15}$-plet fermion 
 with the half-periodic boundary condition 
 has the KK mass spectrum as shown in Eqs.~(\ref{10KKspectrum}) 
 and (\ref{15KKspectrum}). 
Applying the Higgs low energy theorem, 
 their contributions to the Higgs-to-digluon coupling 
 are calculated as 
\bea
C_{gg}^{{\rm KK10}} 
 &\simeq& F(3m_W) + F(2m_W) + 2F(m_W), \\
\label{digluon10}
C_{gg}^{{\rm KK15}} &\simeq& 
 F(4m_W) + F(3m_W) + 2F(2m_W) + 2F(m_W) 
\label{digluon15}
\eea
where the function $F(m_W)$ is given by \cite{MO2}
\bea
F(m_W) &=& \frac{\alpha_{s}}{12\pi v} 
 \sum_{n=1}^\infty \frac{\partial}{\partial \log v} \left[ \log \sqrt{M^2+(m_{n + \frac{1}{2}} + m_W)^2} 
+ \log \sqrt{M^2+(m_{n + \frac{1}{2}} - m_W)^2} \right] \nonumber \\
&\simeq&
-\frac{\alpha_{s}}{6\pi v} \left(\frac{m_W}{M_{\rm KK}} \right)^2
\sum_{n=0}^\infty 
\frac{\left( n+\frac{1}{2} \right)^2- c_B^2}
{\left( \left(n+\frac{1}{2}\right)^2+ c_B^2 \right)^2} 
=
-\frac{\alpha_{s}}{12 \pi v} \left(\frac{m_W}{M_{\rm KK}} \right)^2
 \frac{\pi^2}{\cosh (\pi c_B)}. 
\eea
Here we have used the approximation $m_W^2 \ll M_{\rm KK}^2$.

As pointed out in \cite{MO}, 
 the KK mode contribution to the Higgs-to-digluon coupling 
 is destructive to the SM one. 
This result originates from the mass splitting 
 between $m_{n,t}^{(+)}$ and $m_{n,t}^{(-)}$ 
 in Eq.~(\ref{KKtopmass})
 or between $m_{n,q}^{(+)}$ and $m_{n,q}^{(-)}$ 
 in Eqs.~(\ref{10KKspectrum}) and (\ref{15KKspectrum}), 
 which are generated by the electroweak symmetry breaking.\footnote{ 
 This mass splitting also plays a crucial role to 
 make the KK-mode loop corrections finite.  
 It has been shown \cite{Maru} that this finiteness is valid 
 for GHU in various space-time dimensions and at any perturbative 
 level. }
This destructive contribution is a typical feature 
 of the GHU, in sharp contrast with the one 
 in the universal extra dimension models \cite{UED}.

\begin{figure}[htbp]
  \begin{center}
   \includegraphics[width=75mm]{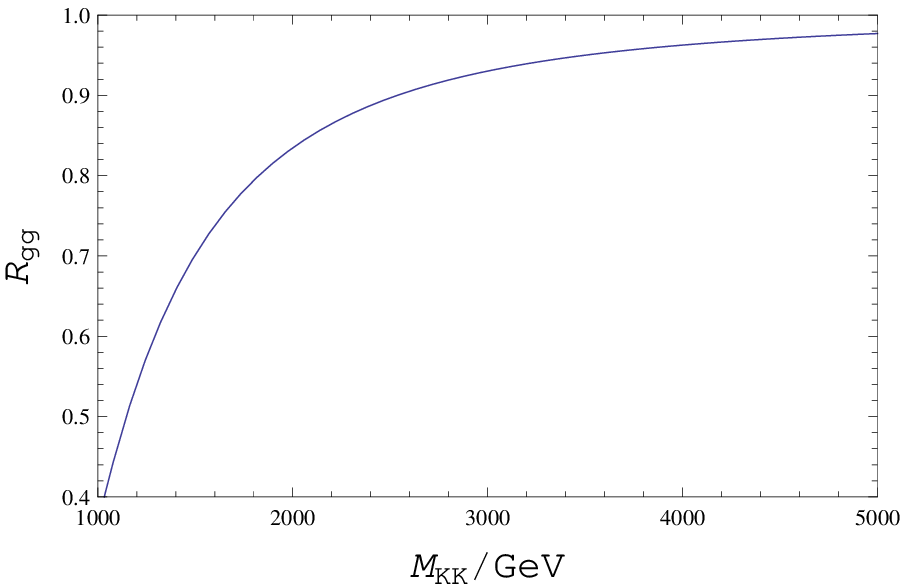}
   \hspace*{5mm}
   \includegraphics[width=75mm]{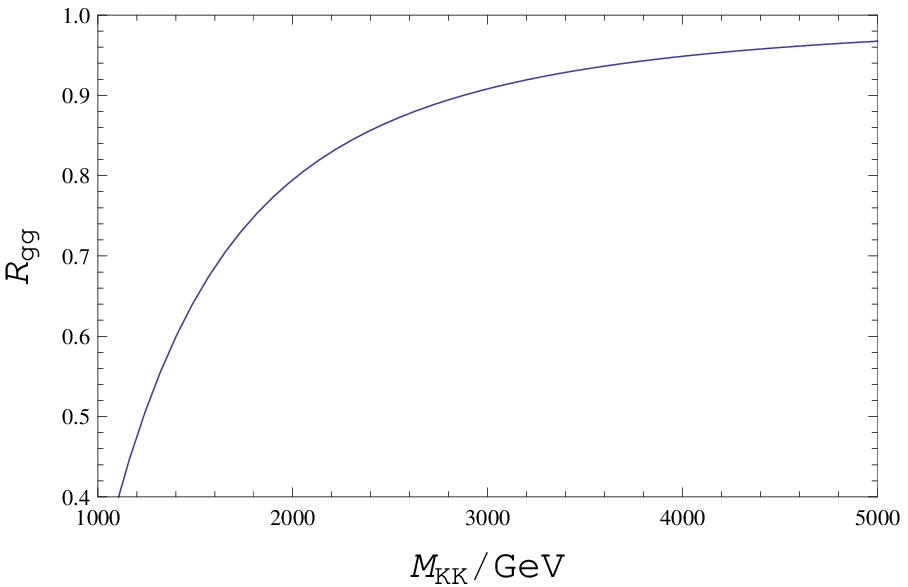}
   \end{center}
  \caption{
The ratio of the Higgs production cross section 
 in our model to the SM one as a function of 
 the compactification scale $M_{\rm KK}$. 
The left (right) panel corresponds to the ${\bf 10}({\bf 15})$-plet case.
The bulk mass $c_B$ is fixed as a function of $M_{\rm KK}$ 
 (along the lines in Fig.~\ref{M0vsMKK}) 
 so as to reproduce $m_h=125$ GeV. 
}
  \label{ggh}
\end{figure}

Now we evaluate the ratio of the Higgs production cross section 
 through the gluon fusion at the LHC 
 in our model to the SM one as 
\bea
R_{gg} &\equiv& 
 \left( 1 + \frac{C_{gg}^{\rm KKtop}+C_{gg}^{\rm KK10(15)}}{C_{gg}^{\rm SMtop}} \right)^2 \nonumber \\
&\simeq&
 \left\{
 \begin{array}{l}
 \left( 1 - \frac{\pi^2}{3} \left(\frac{2 m_W}{M_{\rm KK}}\right)^2 
- \left( \frac{m_W}{M_{\rm KK}} \right)^2 \frac{15\pi^2}{\cosh(\pi c_B)} \right)^2 {\rm for}~{\bf 10}{\rm  -plet.} \\
  \left( 1 - \frac{\pi^2}{3} \left(\frac{2 m_W}{M_{\rm KK}}\right)^2 -
\left( \frac{m_W}{M_{\rm KK}} \right)^2 \frac{25\pi^2}{\cosh(\pi c_B)} \right)^2 {\rm for}~{\bf 15}{\rm  -plet.} \\
  \end{array}
 \right.
\label{R_g}
\eea
As we have investigated in the previous section, 
 in order to reproduce $m_h=125$ GeV 
 there is a relation between $M_{\rm KK}$ and 
 $m_0^{(\pm)}$ (or equivalently, $c_B$) 
 shown in Fig.~\ref{M0vsMKK}. 
Thus, the ratio $R_{gg}$ is described as a function of 
 only the compactification scale $M_{\rm KK}$. 
The results are shown in Fig.~\ref{ggh}.

Since the KK mode contribution to the Higgs production cross section 
 is destructive to the SM process, 
 the resultant cross section becomes smaller 
 than the SM value as the compactification scale is lowered. 
The Higgs boson properties observed at the LHC experiments 
 are mostly consistent with the SM expectations, 
 so that the LHC data can provide a lower bound on  
 the compactification scale not to make the Higgs boson 
 production rate too small. 
The Higgs boson has couplings with the weak gauge bosons 
 and SM fermions at the tree level, which the KK mode 
 contribution to are negligible. 
Thus, the signal strength for the process 
 $gg \to h \to Z Z^*$, for example, is altered essentially 
 by a change of the Higgs production cross section 
 through the gluon fusion process. 
Although the current LHC result for the Higgs boson signal 
 still allow a large deviation from the SM prediction, 
 we here consider a constraint on the Higgs production cross section 
 to be larger than 90 (95) \% of the SM model prediction, 
 $R_{gg} \geq 0.9$ ($R_{gg} \geq 0.95$). 
 From Figs.~\ref{ggh} and \ref{M0vsMKK}, we can read off 
 the lower bound on $M_{\rm KK}$ and $m_0^{(\pm)}$ 
 for the case of the ${\bf 10}$ and ${\bf 15}$-plets, respectively. 
With the lower bound on $M_{\rm KK}$ and $m_0^{(\pm)}$, 
 we can calculate the lightest KK mode mass 
 from the mass spectrum formulas in Eqs.~(\ref{10KKspectrum}) 
 and (\ref{15KKspectrum}). 
For the case of the ${\bf 10}$-plet, 
 we have found the lightest KK mode mass as 
 1.9 TeV (2.4 TeV) for $R_{gg} =0.9$ ($R_{gg}=0.95$), 
 while 2.5 TeV (2.7 TeV) for $R_{gg} =0.9$ ($R_{gg}=0.95$), 
 for the case of the ${\bf 15}$-plet.
The results are shown in Table~\ref{table1}. 
The lower bounds are found to be at TeV, 
 so that such exotic colored particles can be discovered 
 at the LHC Run II with $\sqrt{s}=13-14$ TeV.

\begin{table}[htb]
  \begin{center}
    \begin{tabular}{|c|c|c|} 
  \hline
  ${\bf 10}$-plet & $R_{gg}=0.9$  & $R_{gg}=0.95$ \\ 
\hline
  $M_{\rm KK}$ (TeV)           & 2.54 &  3.45  \\
  $m_0^{(\pm)}$ (TeV)          & 2.05 &  2.91  \\
  $m_{{\rm lightest}}$ (TeV)   & 1.91 &  2.77  \\
\hline 
\hline 
  ${\bf 15}$-plet & $R_{gg}=0.9$  & $R_{gg}=0.95$ \\ 
\hline
  $M_{\rm KK}$ (TeV)           & 2.88 &  4.05  \\
  $m_0^{(\pm)}$ (TeV)          & 2.73 &  3.87  \\
  $m_{{\rm lightest}}$ (TeV)   & 2.57 &  3.71  \\
\hline
\end{tabular}
  \caption{
The lower bound on the KK mode masses 
 from $0.9 \leq R_{gg}$ and $0.95 \leq R_{gg}$ 
 for the cases with the ${\bf 10}$ and ${\bf 15}$-plets. 
}
  \label{table1}
  \end{center}
\end{table}

\subsection{Higgs decay to diphoton}
Next we calculate the KK model contributions 
 to the Higgs-to-diphoton coupling of the dimension five operator, 
\bea
{\cal L}_{{\rm eff}} = C_{\gamma\gamma} h F_{\mu\nu} F^{\mu\nu},  
\eea
 where $F_{\mu\nu}$ denotes the photon field strength. 
The Higgs low energy theorem also allows us  
 to extract the coefficient from the 1-loop RGE 
 of the QED coupling. 
In the SM, the diphoton coupling is induced by 
 the top quark and $W$-boson loop corrections. 
In addition to it, 
 we have the contributions from  the KK modes of 
 top quark, $W$-boson, and the ${\bf 10}$-plet 
 or the ${\bf 15}$-plet.

We begin with the SM top loop contribution.  
As a good approximation, we have 
\bea
C_{\gamma\gamma}^{{\rm SMtop}} \simeq 
 \frac{e^2 b_1^t}{24\pi^2 v} \frac{\partial}{\partial \log v} \log m_t 
= \frac{2\alpha_{em}}{9\pi v}, 
\label{SMt2gamma}
\eea
 where $b_1=(2/3)^2 \times 3 =4/3$ is a top quark contribution 
 to the QED beta function coefficient, 
 and $\alpha_{em}$ is the fine structure constant. 
Corresponding KK top quark contribution is given by
\bea
C_{\gamma\gamma}^{{\rm KKtop}} &\simeq& \frac{e^2 b_1^t}{24\pi^2 v} 
\sum_{n=1}^\infty \frac{\partial}{\partial \log v} \left[ \log (m_n + m_t)  + \log (m_n - m_t) \right] \nonumber \\
&\simeq& 
-\frac{2\alpha_{em}}{9\pi v} \times \frac{\pi^2}{3} 
 \left(\frac{2 m_W}{M_{\rm KK}} \right)^2. 
\label{KKt2gamma}
\eea
Similar to the Higgs-to-digluon coupling, 
 the KK top contribution is destructive 
 to the SM top contribution.

The SM $W$-boson loop contribution is calculated as 
\bea
C_{\gamma\gamma}^W &\simeq& \frac{e^2}{32\pi^2 v} b_1^W 
 \frac{\partial}{\partial \log v} \log m_W = -\frac{7\alpha_{em}}{8\pi v}
\label{SMW2gamma}
\eea
 where $m_W=g_2 v/2$, and $b_1^W=-7$ is a $W$-boson contribution 
 to the QED beta function coefficient. 
This is a rough estimation of the $W$-boson loop contributions 
 since $4 m_W^2/m_h^2 \gg 1$ is not well satisfied. 
For our numerical analysis in the following, 
 we actually use the known loop functions 
 for the top quark and $W$-boson loop corrections.

In our model, the KK mode mass spectrum of the $W$-boson 
 is given by 
\bea
 m_{n,W}^{(\pm)} = m_n \pm m_W,  
\eea  
 so that the contribution from KK $W$-boson loop diagrams
 is found to be 
\bea
C_{\gamma\gamma}^{{\rm KKW}} &=& 
 \frac{e^2}{32\pi^2 v} b_1^W \sum_{n=1}^\infty 
 \frac{\partial}{\partial \log v} 
\left[ \log (m_n + m_W) + \log (m_n - m_W) \right]  \nonumber \\
&\simeq& 
\frac{7\alpha_{em}}{8 \pi v} \frac{\pi^2}{3} 
 \left( \frac{m_W}{M_{\rm KK}} \right)^2. 
\label{KKW2gamma}
\eea
Note again that the KK $W$-boson contribution 
 is destructive to the SM $W$-boson contribution.

Finally, the ${\bf 10}$-plet or ${\bf 15}$-plet loop contributions 
 can be read from the KK-mode mass spectrum 
 in Eqs.~(\ref{10KKspectrum}) or (\ref{15KKspectrum}) 
 and the electric charges of each modes: 
\bea
C_{\gamma\gamma}^{{\rm KK10}} 
 &\simeq& (Q-1)^2 \hat{F}(3m_W) + (Q-1)^2 \hat{F}(m_W) + Q^2 \hat{F}(2m_W) + (Q+1)^2 \hat{F}(m_W), \\
\label{diphoton10}
C_{\gamma\gamma}^{{\rm KK15}} &\simeq& 
 (Q-4/3)^2 \hat{F}(4m_W) + (Q-4/3)^2 \hat{F}(2m_W) + (Q-1/3)^2 \hat{F}(3m_W) \nonumber \\
 &&+ (Q-1/3)^2 \hat{F}(m_W) + (Q+2/3)^2 \hat{F}(2m_W) + (Q+5/3)^2 \hat{F}(m_W)
\label{diphoton15}
\eea
where $Q$ is a $U(1)^\prime$ charge for 
 the ${\bf 10}$ and ${\bf 15}$-plets, and 
\bea 
{\hat F}(m_W) \simeq 
 -\frac{\alpha_{em}}{2 \pi v} \left(\frac{m_W}{M_{\rm KK}} \right)^2
 \frac{\pi^2}{\cosh (\pi c_B)}. 
\eea

Putting all together, 
 we find the ratio of the partial decay width of $h \to \gamma \gamma$ 
 in our model to the SM one as 
\bea
R_{\gamma\gamma} &\equiv& \left( 1+ \frac{C_{\gamma\gamma}^{{\rm KKtop}} 
+ C_{\gamma\gamma}^{{\rm KKW}} + C_{\gamma\gamma}^{{\rm KK10(15)}}}
{C_{\gamma\gamma}^{{\rm SMtop}} + C_{\gamma\gamma}^W} \right)^2
\nonumber \\
&\simeq& \left\{
\begin{array}{l}
\left( 
 1 + \frac{\pi^2}{141} \left(\frac{m_W}{M_{\rm KK}} \right)^2 
 +  \left[ (3^2+1^2)(Q-1)^2 +4Q^2 +(Q+1)^2 \right] 
  \left(\frac{m_W}{M_{\rm KK}} \right)^2 \frac{36 \pi^2}{47 \cosh(\pi c_B)}
  \right)^2 \\
  \hspace*{11.5cm} {\rm for}~{\bf 10}{\rm -plet}. \\
 \left( 
 1 + \frac{\pi^2}{141} \left(\frac{m_W}{M_{\rm KK}} \right)^2 
 + \left[ (4^2+2^2) \left( Q-\frac{4}{3} \right)^2 + (3^2+1^2) \left( Q-\frac{1}{3} \right)^2 \right. \right. \\
 \hspace*{2.5cm}
 + \left. \left. 2^2 \left( Q + \frac{2}{3} \right)^2 + \left( Q + \frac{5}{3} \right)^2
 \right] 
  \left(\frac{m_W}{M_{\rm KK}} \right)^2 \frac{36 \pi^2}{47 \cosh(\pi c_B)}
  \right)^2~{\rm for}~{\bf 15}{\rm -plet}. \\
 \end{array}
 \right. 
 \nonumber \\
\eea
%
For a fixed $Q$, the KK mode contributions becomes larger 
 as $M_{\rm KK}$ goes down. 
As expected, a large $|Q| \gg 1$ significantly alters 
 $R_{\gamma\gamma}$ from $1$, for a fixed $M_{\rm KK}$.

\begin{figure}[ht]
  \begin{center}
   \includegraphics[width=75mm]{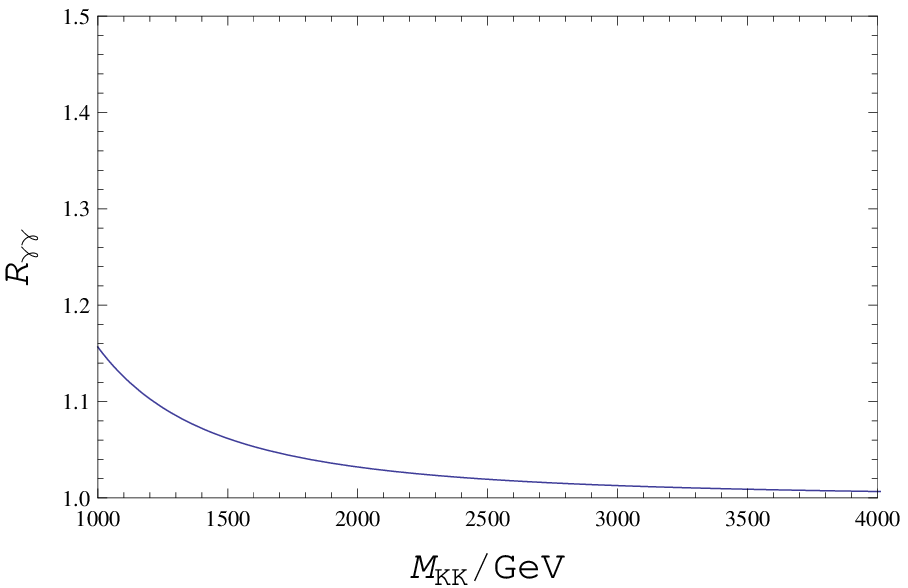}
   \hspace*{5mm}
   \includegraphics[width=75mm]{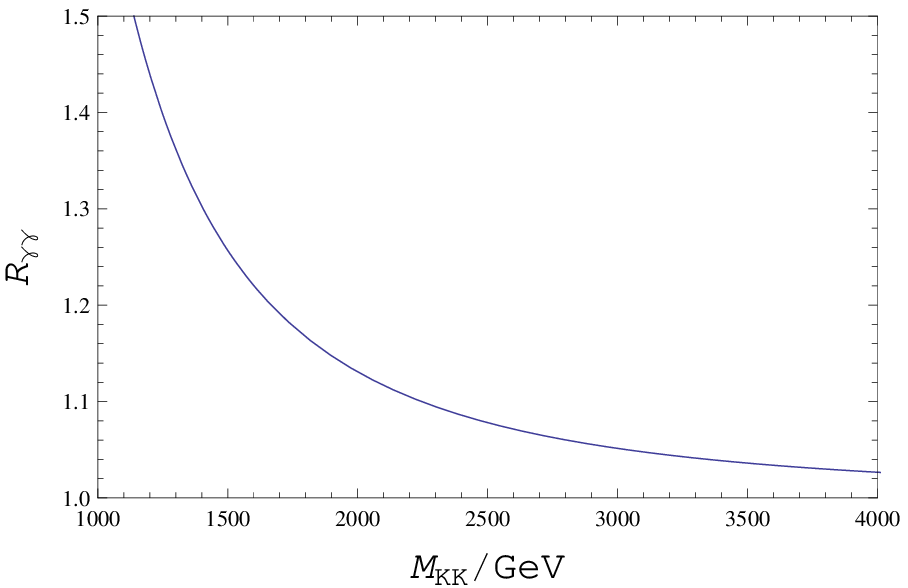}\\
   \vspace*{3mm}
   \includegraphics[width=75mm]{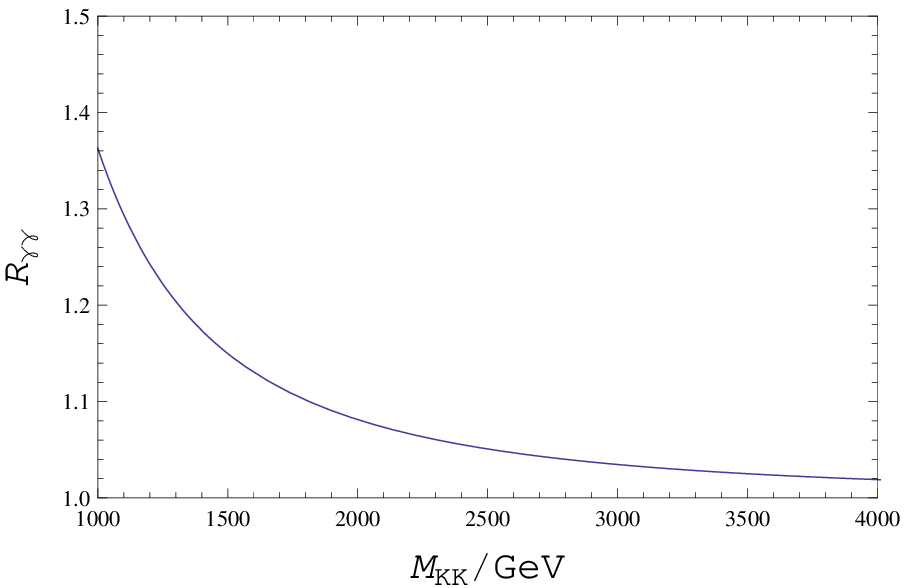}
   \hspace*{5mm}
   \includegraphics[width=75mm]{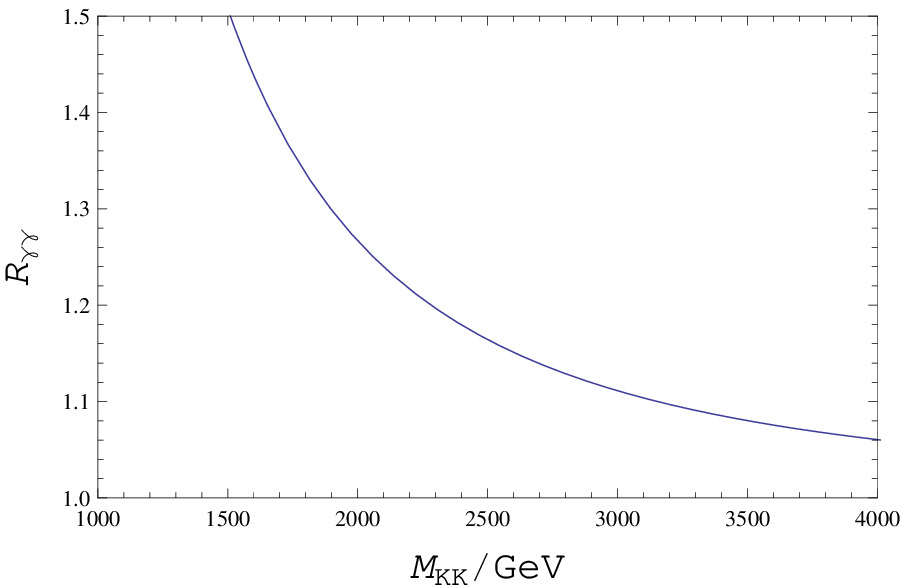}
   \end{center}
  \caption{
The ratio $R_{\gamma \gamma}$ 
 for the ${\bf 10}$-plet case (upper panels) 
 and the ${\bf 15}$-plet case (lower panels) 
 as a function of $M_{\rm KK}$ 
 with the relations between the bulk mass and $M_{\rm KK}$ 
 depicted in Fig.~\ref{M0vsMKK}. 
We have fixed $Q=2/3$ ($Q=5/3$) for the upper-left (upper-right) panel 
 in the ${\bf 10}$-plet case (upper panels), 
 while $Q=1$ ($Q=2$) for the lower-left (lower-right) panel 
 in the ${\bf 15}$-plet case.
}
  \label{diphoton}
\end{figure}

Although the $U(1)^\prime$ charge $Q$ is a free parameter 
 of the model, we have phenomenologically favored values 
 for it from the following discussion. 
As discussed in \cite{GHDM} (see also \cite{GHDM2}),
 the lightest KK mode of the half-periodic bulk fermion, 
 independently of the background metric, is stable 
 in the effective 4-dimensional theory 
 due to the accidental $Z_2$ discrete symmetry. 
However, a colored stable particle is cosmologically disfavored. 
An easy way to solve this phenomenological problem 
 is to introduce a mixing between the lightest colored KK fermion 
 and a SM quark on the brane, so that the lightest KK fermion 
 can decay to the SM quarks. 
There are two choices for the $U(1)'$ charge to make 
 the electric charge of the lightest KK mode 
 to be $-1/3$ or $2/3$ for realizing a mixing 
 with either the down-type quarks or up-type quarks. 
For the ${\bf 10}$-plet case we choose $Q=2/3$ or $5/3$, 
 while $Q=1$ or $2$ for the ${\bf 15}$-plet case. 
Fig.~\ref{diphoton} shows the results for $R_{\gamma \gamma}$ 
 as a function of $M_{\rm KK}$ for the ${\bf 10}$ and ${\bf 15}$-plet 
 cases with the choices, $Q=2/3$ or $5/3$ and  $Q=1$ or $2$, 
 respectively. 
Here the bulk mass terms are fixed along the lines 
 in Fig.~\ref{M0vsMKK}  so as to give $m_h=125$ GeV. 

\subsection{$gg \to h \to \gamma\gamma$}
Let us finally calculate the ratio of 
 the signal strength of the process 
 $gg \to h \to \gamma \gamma$ in our model to the SM one. 
For the ${\bf 10}$-plet, we have 
\bea
{\rm R} &\equiv& \frac{\sigma(gg \to h \to \gamma \gamma)
}{\sigma(gg \to h \to \gamma \gamma)_{\rm SM}}
=R_{gg} \times R_{\gamma\gamma} \nonumber \\
&\simeq& 1 - \frac{374}{141} \pi^2  \left(\frac{m_W}{M_{\rm KK}} \right)^2  \nonumber \\
&&+ \left[ -30 + \frac{72}{47} \left\{ 10(Q-1)^2 + 4Q^2 + (Q+1)^2 \right\} \right] 
\left(\frac{m_W}{M_{\rm KK}} \right)^2 \frac{\pi^2}{\cosh(\pi c_B)}, 
\eea
while for the ${\bf 15}$-plet  
\bea
{\rm R} &\simeq& 1 - \frac{374}{141} \pi^2  \left(\frac{m_W}{M_{\rm KK}} \right)^2  
+ \left[ -50 + \frac{72}{47} \left\{ 20 \left( Q-\frac{4}{3} \right)^2 \right. \right. \nonumber \\
&& \left. \left. + 10 \left( Q-\frac{1}{3} \right)^2 
+ 4\left( Q + \frac{2}{3} \right)^2 + \left( Q+\frac{5}{3} \right)^2 \right\} \right] 
\left(\frac{m_W}{M_{\rm KK}} \right)^2 \frac{\pi^2}{\cosh(\pi c_B)}.  
\eea

\begin{figure}[ht]
  \begin{center}
   \includegraphics[width=75mm]{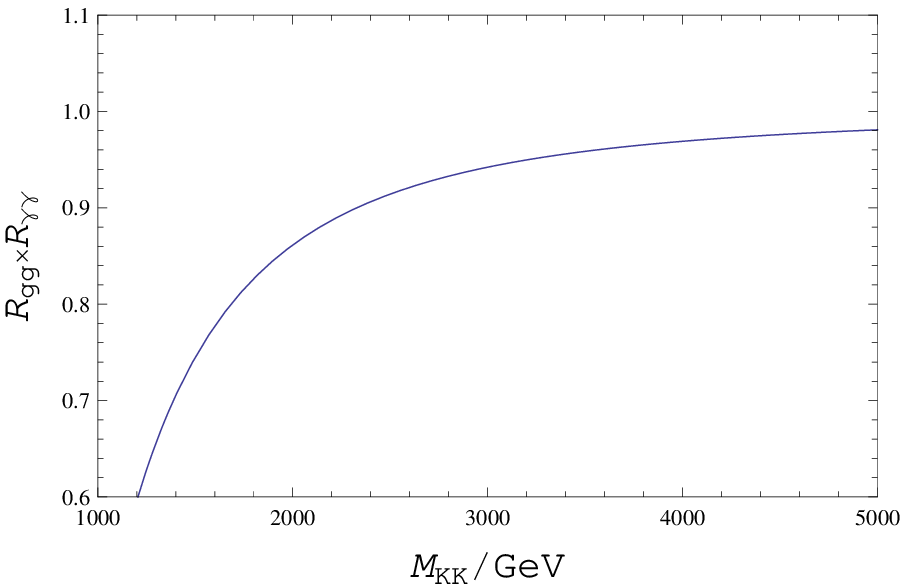}
   \hspace*{5mm}
   \includegraphics[width=75mm]{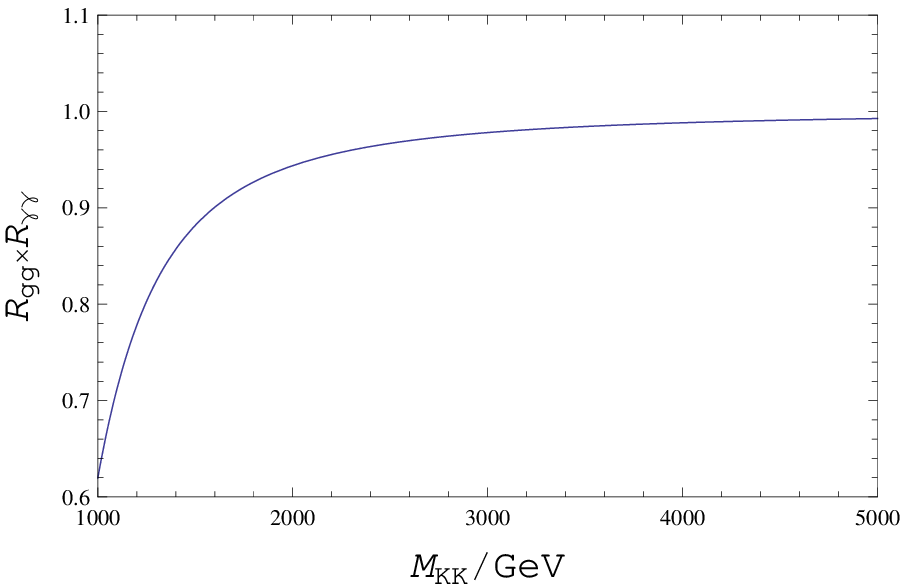} \\
   \vspace*{3mm}
    \includegraphics[width=75mm]{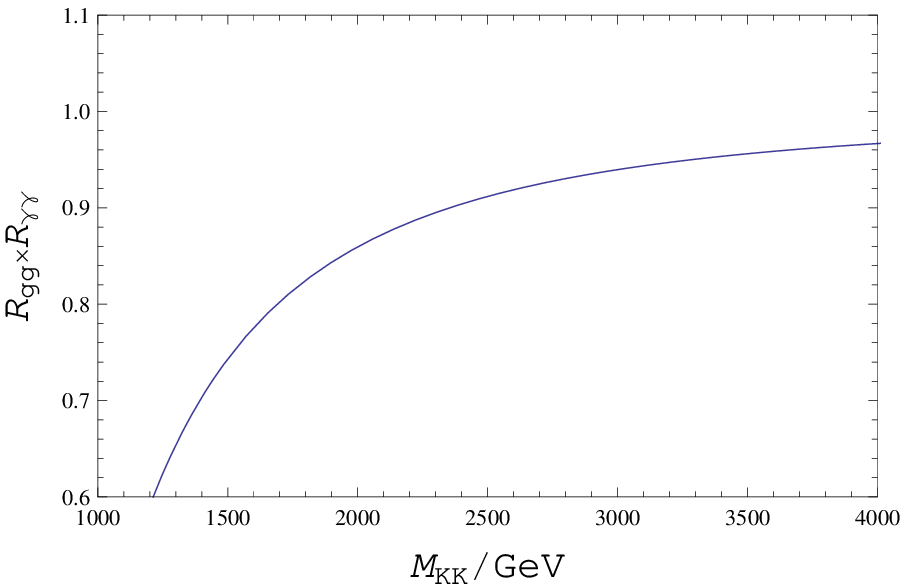}
   \hspace*{5mm}
   \includegraphics[width=75mm]{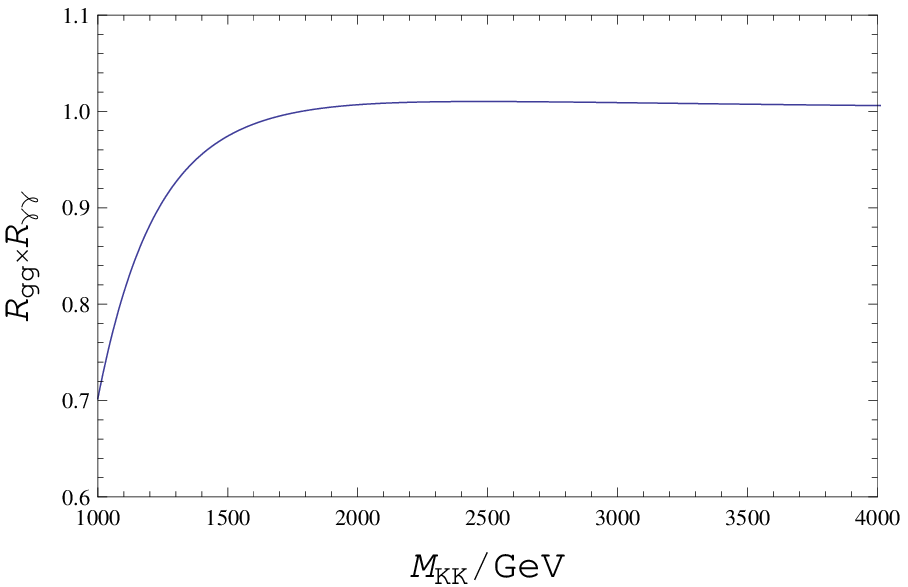}
   \end{center}
  \caption{
The diphoton signal strength normalized by the SM prediction 
 in the GHU model with the ${\bf 10}$-plet bulk fermions (upper panels) 
 and the ${\bf 15}$-plet bulk fermions (lower panels) 
 as a function of $M_{\rm KK}$. 
The $U(1)'$ charge of the lightest KK mode fermion 
 in the ${\bf 10}$-plet is fixed to be $Q=2/3$ ($Q=5/3$) 
 for the upper-left (upper-right) panel.
For the ${\bf 15}$-plet case, we have fixed $Q=1$ ($Q=2$)
 for the lower-left (lower-right) panel.
}
  \label{gghdiphoton}
\end{figure}

For the two cases, we plot the ratio R as a function of  
 the compactification scale $M_{\rm KK}$ in Fig.~\ref{gghdiphoton}. 
The upper two panels corresponds to the ${\bf 10}$-plet case, 
 where we have fixed $Q=2/3$ ($Q=5/3$) in the upper-left (upper-right) panel. 
The lower two panels corresponds to the ${\bf 15}$-plet case, 
 where we have fixed $Q=1$ ($Q=2$) in the lower-left (lower-right) panel. 
Comparing Fig.~\ref{gghdiphoton} with Fig.~\ref{ggh}, 
 we can see that the ratio R is mainly controlled 
 by the Higgs production cross section in Fig.~\ref{ggh},  
 and the signal strength of the diphoton events  
 is smaller than the SM expectation.

\section{KK mode fermions at the LHC}
Since the mass of the lightest KK mode fermions 
 of the ${\bf 10}$ and ${\bf 15}$-plets  
 can be as low as a few TeV, 
 the colored fermions are accessible 
 to the LHC run II with the upgraded collider energy 
 $\sqrt{s}=13-14$ TeV. 
We have assigned the $U(1)^\prime$ charge 
 so as for the lightest KK mode fermions 
 to have the electric charge of either $-1/3$ or $2/3$ 
 and hence to decay to the SM quarks. 
When the lightest KK mode fermions couple 
 mainly with the third generation quarks, 
 the LHC phenomenology for them is similar 
 to the one for the 4th generation quarks. 
The KK mode fermions, once produced 
 dominantly through the gluon fusion process, 
 decay to the $W$-boson/$Z$-boson/Higgs boson 
 and top/bottom quark through the charged/neutral current. 
See, for example, \cite{4th} for the current limit 
 on the heavy quarks at the LHC.

\begin{figure}[htbp]
  \begin{center}
 \includegraphics[width=50mm, angle=90]{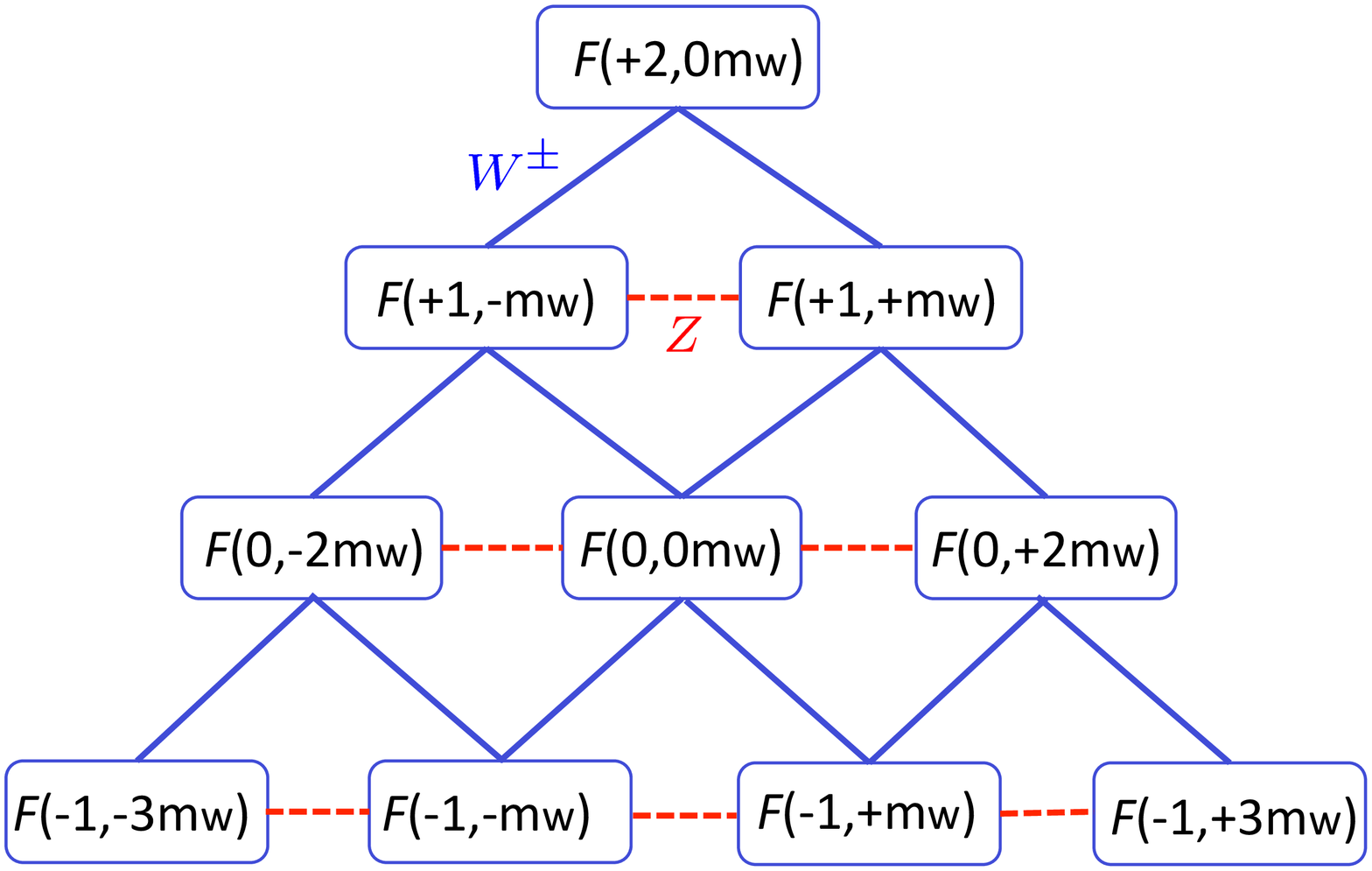}
   \end{center}
  \caption{
A sketch of the interactions among the mass eigenstates 
 for the ${\bf 10}$-plet case.  
The diagonal (solid) lines represent the interactions 
 between the connected two mass eigenstates and the $W$-boson, 
 while the horizontal (dashed) lines represent the interactions 
 between the connected two mass eigenstates and the $Z$-boson. 
Here the symbol $F(q,\alpha m_W)$ denotes the mass eigenstate  
 corresponding to the mass eigenvalue in Eq.~(\ref{10KKspectrum}). 
For example, the mass eigenstate $F(-1, 3 m_W)$ 
 has the mass squared eigenvalue of
 $( m_{\frac{1}{2}}+3 m_W )^2+M^2$. 
}
  \label{cascade}
\end{figure}

Note that in our model the bulk fermions has been introduced 
 as the ${\bf 10}$-plet or the ${\bf 15}$-plet, 
 and 10 or 15 fermion mass eigenstates appear 
 after the electroweak symmetry breaking, 
 as described in Eqs.~(\ref{10KKspectrum}) and (\ref{15KKspectrum}). 
Therefore, if $m_0^{(\pm)}$ is low enough, say, $\leq 3$ TeV 
 heavy colored particles with a variety of electromagnetic 
 charges can be produced at the LHC run II. 
Since the mass eigenstates couples with each other 
 through the weak gauge bosons, a heavy fermion, 
 once produced at the LHC, causes cascade decays 
 to lighter mass eigenstates and the weak gauge boson, 
 which end up with the lightest KK mode fermion, 
 followed by the decay to the SM quark 
 through the charged/neutral currents. 
The existence of the variety of fermion mass eigenstates 
 and their cascade decay once produced at the LHC 
 are characteristic feature of our GHU model. 
Fig.~\ref{cascade} sketches the interactions 
 among the various mass eigenstates for the ${\bf 10}$-plet case. 
In the figure, we can see a characteristic structure 
 of the interactions among the mass eigenstates and 
 the $Z$-boson, namely, the $Z$-boson only couples 
 with two different mass eigenstates. 
As has been pointed out in \cite{MO3}, 
 there is no KK mode contribution to the effective Higgs-to-$Z \gamma$
 in the GHU, because of this characteristic coupling manner. 

\vspace*{5mm}

\section{Conclusions}
In this paper, we have performed the Higgs boson mass analysis 
 in a 5-dimensional $SU(3)\times U(1)'$ GHU model. 
In order to obtain the 125 GeV Higgs mass, 
 we have extended a simple GHU model by introducing colored bulk fermions 
 with a half-periodic boundary condition and bulk masses. 
For concreteness, we have considered the fermions 
 in the ${\bf 6}$, ${\bf 8}$, ${\bf 10}$ and ${\bf 15}$-plets of $SU(3)$ 
 with a $U(1)'$ charge $Q$. 
Employing the gauge-Higgs condition, we have shown 
 in the RGE analysis that the 125 GeV Higgs boson mass 
 can be realized in the presence of 
 the half-periodic bulk fermions by adjusting their bulk masses. 
Our interest is how the bulk masses can be lowered 
 so that the lightest KK mode fermion might be detectable 
 by the LHC run II. 
In this regard, the cases with the ${\bf 6}$ and ${\bf 8}$-plets 
 are not attractive since the lightest KK mode masses 
 are required to be $\gtrsim 4$ TeV 
 in order to reproduce the 125 GeV Higgs boson mass. 
On the other hand, the lightest KK mode masses 
 being even less than 1 TeV can achieve $m_h=125$ GeV 
 in the ${\bf 10}$ and ${\bf 15}$-plet cases. 
Thus, we have focused on these two cases.

The KK mode fermions of the ${\bf 10}$ and ${\bf 15}$-plets  
 have contributions to the Higgs-to-digluon and
 Higgs-to-diphoton couplings. 
We have evaluated the contributions and found that 
 the Higgs boson production cross section is reduced 
 as the mass of the lightest KK mode fermions becomes lighter. 
Thus, the LHC results can be used to derive the lower bound on 
 the lightest KK mode mass to be consistent 
 with the LHC data. 
When $5-10$ \% reduction of the production cross section 
 is required, we have found the lower bound at a few TeV, 
 reproducing the 125 GeV Higgs boson mass.  
The colored KK mode fermions with a few TeV masses 
 are accessible to the LHC run II. 
The $U(1)'$ charges for the lightest KK mode fermions  
 are assigned to be either $-1/3$ or $2/3$,  
 so that they can decay to the SM quarks 
 through the mass mixing introduced on the brane. 
For the charge assignment, 
 the KK mode fermion contributions to the signal strength 
 of the Higgs-to-diphoton evens are mainly controlled 
 by the Higgs production cross section. 
LHC phenomenology for the KK mode fermions are 
 similar to the one for the 4th generation heavy quarks, 
 but our model includes more exotic quarks 
 with a variety of masses and electric charges, 
 which originates form the ${\bf 10}$ or ${\bf 15}$-plets. 
Once a heavy KK mode fermion is produced at the LHC, 
 it decays to lighter mass eigenstates. 
This cascade decay ends up with the lightest KK mode fermion, 
 followed by its decay to the SM quark and 
 the weak boson/Higgs boson. 
Search for the KK mode fermions at the LHC run II 
 is an interesting topic and 
 we leave it for future work.

\section*{Acknowledgments}
The work of N.M. is supported in part by the Grant-in-Aid 
 for Scientific Research from the Ministry of Education, 
 Science and Culture, Japan No. 24540283.
The work of N.O. is supported in part 
 by the DOE Grant No. DE-FG02-10ER41714.


\end{document}